\documentclass{tMPH2e}

\usepackage{graphicx,wasysym}
\usepackage{dcolumn}
\usepackage{bm}
\usepackage{natbib}
\usepackage{tabulary}

\newcommand{\Tm}{T_{\mathrm{m}}}

\newcommand{\Ts}{T_{\mathrm{s}}}
\newcommand{\Th}{T_{\mathrm{h}}}
\newcommand{\To}{T_{\mathrm{o}}}

\begin{document}

\title{Time scales of supercooled water and implications for reversible polyamorphism\\  }

\author{David T. Limmer$^{a}$
\vspace{6pt} 
and David Chandler$^{b}$$^{\ast}$\thanks{$^\ast$Corresponding author. Email: chandler@berkeley.edu}
\\\vspace{6pt}  $^{a}${\em{Princeton Center for Theoretical Science, Princeton University, Princeton, NJ 08544 USA}};
$^{b}${\em{Department of Chemistry, University of California, Berkeley, Berkeley CA 94720 USA}}\\\vspace{6pt} \received{\today}}

\maketitle

\begin{abstract}
Deeply supercooled water exhibits complex dynamics with large density fluctuations, ice coarsening and characteristic time scales extending from picoseconds to milliseconds.  Here, we discuss implications of these time scales as they pertain to two-phase coexistence and to molecular simulations of supercooled water. Specifically, we argue that it is possible to discount liquid-liquid criticality because the time scales imply that correlation lengths for such behavior would be bounded by no more than a few nanometers.  Similarly, it is possible to discount two-liquid coexistence because the time scales imply a bounded interfacial free energy that cannot grow in proportion to a macroscopic surface area.  From time scales alone, therefore, we see that coexisting domains of differing density in supercooled water can be no more than nano-scale transient fluctuations. \bigskip

\begin{keywords} coarsening, nucleation, relaxation, supercooled,  water
\end{keywords}\bigskip
\end{abstract}

\section{Introduction}
For more than two decades, it has been suggested that anomalous properties of liquid water reflect two distinct liquids and a low-temperature critical point at supercooled conditions~\cite{Poole1992}. Yet Binder has observed~\cite{Binder2014} that two-liquid criticality defined in terms of a divergent length scale is impossible at deeply supercooled conditions.  Specifically, growing lengths coincide with growing equilibration times, and the time available to equilibrate can be no longer than the time it takes the metastable liquid to crystallize.  In other words, metastability or instability implies an upper bound to the size of fluctuations that can relax in the liquid.  For water, we argue, the bound seems to be no larger than 2 or 3 nm, corresponding to volumes containing fewer than 1000 molecules.
  
This bound is fundamentally different than a cutoff imposed by the practicality of a finite simulation cell.  Transient fluctuations on smaller length scales might seem interpretable in terms of something like a liquid-liquid transition, but the bound implies one can never reach large enough scales to know if that interpretation is correct.  The interpretation certainly seems unnecessary because reasonable molecular models known to not exhibit two-liquid behavior and general arguments independent of molecular forcefield do account for equilibrium anomalies of water~\cite{Molinero, Limmer2011, Patey2013}  and nonequilibrium amorphous ices~\cite{Limmer2014a,Limmer2014}. 

We will see that the principal governing parameter is the ratio of liquid's metastable lifetime, $\tau_\mathrm{MS}$, to its structural relaxation time, $\tau_\mathrm{R}$.  If $\tau_\mathrm{MS} / \tau_\mathrm{R}$ is large enough, conceptions of liquid-liquid coexistence and criticality can be good approximations.  Physically realizable examples include colloidal mixtures \cite{Asherie1996} and protein solutions \cite{Broide1996}.  

While water is not among those examples, some recent simulations of water models are interpreted as demonstrating two-liquid-like behavior \cite{Palmer2014,kesselring2012,poole2013}.  Other works, such as Refs. \cite{Limmer2013} and \cite{English2013}, seem to discount the possibility. The disagreement is not an issue of the force field chosen to model water, as has been noted in Ref.~\cite{Limmer2013}.  Rather, disagreement about two-liquid-like bistability has to do with equilibration or reversibility.  Indeed, Ref.~\cite{Limmer2013} shows that this bistability is reproduced by constraining the distribution of crystal order parameter to that of the standard liquid state, and that this bistability disappears as the distribution is allowed to relax. 

This general result, independent of free energy sampling method, argues that simulations finding two-liquid behavior, for example \cite{Palmer2014,kesselring2012,poole2013} and \cite{Abascal2012}, have failed to control reversibility in the presence of emergent crystal order.  Establishing reversibility in molecular simulations of supercooled water is possible but difficult both because structural relaxation in the supercooled liquid is intrinsically slow (it is, after all, a glass former), and because the time scales required to relax crystal-order fluctuations can be orders of magnitude longer than those required to relax density fluctuations.  (The Appendix illustrates this time-scale separation.) 

Section 2 gets to the root of the issue by providing estimates of time scales for supercooled water, specifically $\tau_\mathrm{R}$ and $\tau_\mathrm{MS}$.  From those estimates, a bound for the correlation length of putative critical fluctuations in water is derived.  Section 3 then argues that interfacial area between domains of different densities in supercooled water are also bounded.  As such, at reversible conditions, these domains are transient and cannot exhibit a well defined surface tension or an interfacial free energy that grows as a fractional power of system size.  The paper concludes Section 4 with a discussion of differences between supercooled water and systems where two-liquid behavior is possible.

Before turning to those points, let us be clear that nothing in the analysis presented below discounts the possibility of more than one \textit{nonequilibrium} amorphous solid.  In fact, estimates discussed in Section 2 make use of tools that can also be used for predicting the emergence of glass transitions and more than one amorphous solid \cite{Limmer2014a, Limmer2014}.  Rather than this \emph{irreversible} polyamorphism, our focus herein is on the possibility of more than one liquid and criticality at \emph{reversible} conditions.

\section{Time scales and length scales}

\subsection{Regimes of supercooled water}

\begin{figure}
\begin{center}
\includegraphics[width=5.5in]{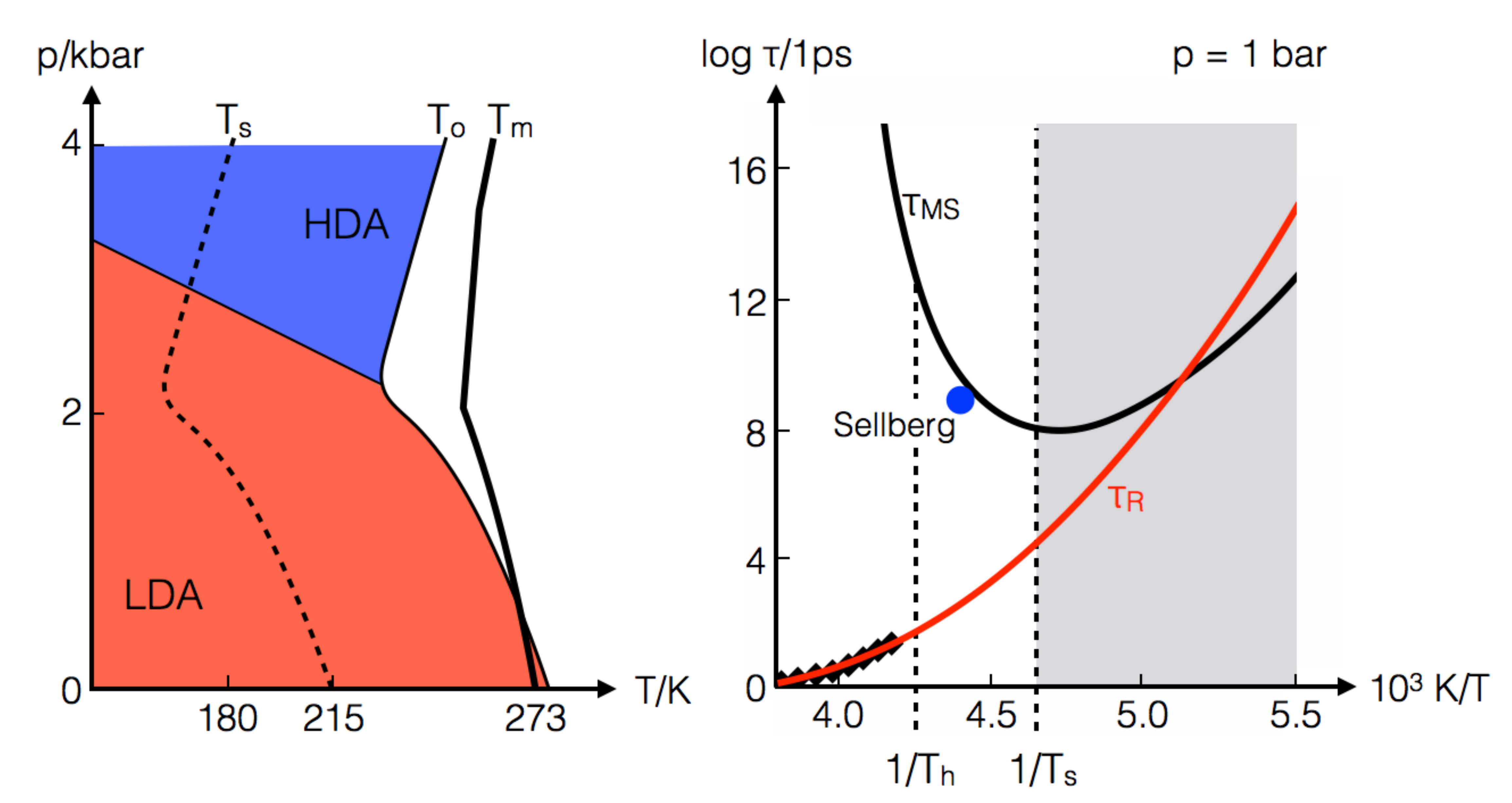}
\caption{(a) Regimes of supercooled water, with $T$ and $p$ denoting temperature and pressure, respectively, and (b) the logarithm of the supercooled liquid's low pressure lifetime, $\tau_\mathrm{MS}$, and structural relaxation time, $\tau_\mathrm{R}$.    Equilibrium liquid-ice coexistence occurs at the melting temperature, $\Tm$, and the crossover to glass forming dynamics occurs in the supercooled liquid below the onset temperature, $\To$.  For the range of supercooled temperatures extending not too far below $\Tm$ or $\To$, the rate-determining step for crystallization is nucleation.
At yet lower temperatures highlighted with grey in Panel (b), $T \lesssim \Ts$, coarsening rather than simple nucleation becomes rate determining. 
($\To$ and $\Ts$ are given in Refs. \cite{Limmer2014} and \cite{Limmer2013a}.)   
The black and red lines in Panel (b) refer to theoretical predictions of $\tau_\mathrm{MS}$ and $\tau_\mathrm{R}$, respectively, i.e., Eqs. (16) and (2) of Ref.~\cite{Limmer2013a}.  The blue circle locates the experimental measurement of metastable lifetime by Sellberg \textit{et al.} \cite{Nilsson2014}.  The temperature $\Th$ (= 235\,K at $p=1$\,atm) is the homogeneous nucleation temperature given by Holten \textit{et al.} \cite{Holten2012}.  It is the highest temperature at which ice crystallization proceeds quickly, i.e., where the metastable lifetime is of order $10^{1\pm 1}$\,s.  The diamonds are experimental results for $\tau_\mathrm{R}$ as reflected in experimental measurements of viscosity \cite{Osipov1977}.}
\label{fig:NewFig1}
\end{center}
\end{figure}

Figure~\ref{fig:NewFig1}(a) shows a phase diagram 
for supercooled water.  The onset temperature, $\To$, is the crossover temperature below which liquid dynamics is heterogeneous and relaxation times grow with decreasing $T$ in a super-Arrhenius fashion. The liquid instability temperature, $T_s$, is the temperature below which nanometer-scale domains of the liquid are unstable.  Fluctuations are thus large for all $T \lesssim \Ts$, but because the time to reorganize the liquid grows rapidly as temperature is lowered, fluctuations observed over finite times can appear largest in the vicinity of $\Ts$.  Not surprisingly, therefore, all estimates of a putative critical temperature for supercooled water are close to $\Ts$. 

Crystallization is avoided and glass is formed when the supercooled liquid is cooled fast enough and to a low enough temperature.  The specific temperature for this transformation depends upon the time scale at which the liquid is driven out of equilibrium \cite{Limmer2014a}.  Given this protocol dependence, it is impossible to illustrate the multitude of possible nonequilibrium phase behaviors in a simple two-dimensional plot.  Nevertheless, one significant feature is captured with the line drawn in the figure between high-density amorphous (HDA) and low-density amorphous (LDA) regions.  

This line marks the $p$ at which the $T$ to reach the nonequilibrium time scale is minimum \cite{Limmer2014}.  At low enough temperatures, this line relates to a first-order-like nonequilibrium transition between HDA and LDA phases.  Observations of the transition show a large range of hysteresis  with the average of the forward and backward transition pressures being close to that line \cite{Loerting2006}.  In the reversible melt near $T=\Ts$, transient mesoscopic domains will necessarily appear either as the first steps in ice coarsening or as precursors to those nonequilibrium amorphous solids.  In a molecular simulation of a small enough system carried out over a small enough time, those transient domains can be confused with two distinct liquids.

For such simulations, locations for the corresponding states illustrated in Fig.~\ref{fig:NewFig1}(a) differ from one model to another~\cite{Limmer2013a}. For example, with the ST2 model of water used in Ref. \cite{Palmer2014}, the corresponding states are shifted to higher temperatures from those of real water by 10 to 15\%.

\subsection{Metastable lifetime and structural relaxation time}
Figure~\ref{fig:NewFig1}(b) shows the temperature variation of $\tau_\mathrm{MS}$ and $\tau_\mathrm{R}$.  The lines are the theoretical predictions of Ref.~\cite{Limmer2013a}. 
At conditions where the liquid persists long enough to make the measurements \cite{Osipov1977}, the theoretical prediction for $\tau_\mathrm{R}$ agrees well with the experimental measurements.  At lower temperatures, $T < \Th \approx 235 $K (i.e., in so-called ``no-man's land''),  agreement with simulations at similar corresponding states gives further confidence in the theory that predicts the red line \cite{Limmer2013a}.  

The theoretical prediction of $\tau_\mathrm{MS}$ agrees with the one experimental measurement of that time for $\Ts < T < \Th$.  The metastable lifetime is a non-monotonic function of temperature because above $\Ts$, nucleation is rate determining in crystal formation, while below $\Ts$, critical nuclei are small and plentiful, and coarsening is rate determining.  Grey shading in Fig.~\ref{fig:NewFig1}(b) highlights the distinction between the two regimes.  The latter regime, with its large fluctuations and coarsening, is the regime of interest when considering possible two-liquid behavior.   

The theoretical curve, essentially an interpolation connecting the two regimes \cite{Limmer2013a}, predicts  
that the shortest lifetime for supercooled water is no longer than $10^{-4}$\,s.  Experimentally, making amorphous ice by cooling requires cooling rates at least as fast as $10^6$\,K/s \cite{kohl2005}.  Based on this cooling rate, dimensional analysis also gives $10^{-4}$\,s as the shortest lifetime, in harmony with the theoretical prediction.

\subsection{Largest length scale of putative critical fluctuations}

The structural relaxation time of the liquid, $\tau_\mathrm{R}$, is the time to equilibrate the liquid on length scale $a$, where $a \approx 0.2$ or 0.3 nm is the characteristic microscopic length of the liquid.  Near presumed criticality, the time to equilibrate over a larger length scale, $\xi$, would be of order $\tau_\xi=\tau_\mathrm{R} (\xi/a)^z$, where $z \approx 3$ \cite{Hohenberg1977}.  But as Binder notes \cite{Binder2014}, $\tau_\xi < \tau_\mathrm{MS}$ because the liquid will not resist crystallization for times longer than $\tau_\mathrm{MS}$.  Accordingly, $\xi/a< (\tau_\mathrm{MS}/\tau_\mathrm{R})^{1/3}$. 

This bound giving the largest correlation length for critical-like fluctuations is applicable in the regime where criticality is imagined to occur, $T \lesssim \Ts$.  Figure~\ref{fig:NewFig1} shows that in this regime at ambient pressures, $\tau_\mathrm{R}$ and $\tau_\mathrm{MS}$ grow with decreasing temperature, and the ratio is $\tau_\mathrm{MS}/\tau_\mathrm{R} \lesssim10^3$ throughout.  Based upon the theoretical predictions, we expect $\tau_\mathrm{MS}/\tau_\mathrm{R} \lesssim10^3$ to remain true at the slightly elevated pressures that are sometimes identified with the putative liquid-liquid critical point, e.g., $ p \approx $ 0.5\,kbar \cite{Holten2012}. 

Specifically, according to the theory \cite{Limmer2013a}, increasing pressure up to 2 kbar will decrease the ratio of $\tau_\mathrm{MS} / \tau_\mathrm{R}$, but by no more than 10 to 20\,$\%$.  The decrease is due to the extent by which increasing pressure increases $ (\partial \tau_\mathrm{R}/ \partial T)_p$ \cite{Limmer2014} and decreases the enthalpy of fusion \cite{Eisenberg2005}. The decrease in enthalpy of fusion causes $\tau_\mathrm{MS}$ to decrease  because the solid-liquid surface tension is proportional to the enthalpy of fusion \cite{Turnbull1950}, a proportionality that has been tested for a molecular model of water \cite{Limmer2012}.   Experimental tests have not yet been performed to test the prediction of $\tau_\mathrm{MS} / \tau_\mathrm{R}$ at elevated pressures. 

Accepting the prediction, as the theory compares favorably with experiment at ambient pressures, it follows that by cooling the system down to $\Ts$, the correlation length of density fluctuations (which may have the physical interpretation of precursor effects of the LDA-HDA nonequilibrium transition at still lower temperatures) can increase by at most one order of magnitude, i.e. $\xi < 2$ or 3 nm.  
Phenomena with such a small largest length scale would seem to be poorly approximated by criticality, which is defined by a diverging correlation length, although in a computer simulation context such a modest increase of the correlation length might be easily mistaken as a signature of a true transition.

\section{Interfaces between finite domains in the metastable liquid}
With a similar argument, one can also conclude that different liquid domains at metastable conditions are bounded in size.  To do so, bear in mind that where nucleation is rate determining, the liquid's lifetime decreases with increasing system size (i.e., the bigger the system the more opportunities for a critical nucleus to appear).  In contrast, in deeply supercooled conditions, where coarsening is rate determining, the metastable lifetime increases with increasing system size (i.e., the bigger the system the more likely to have misaligned crystal domains).  Dependence on system size in the latter case is weak, growing as a fractional power \cite{Avrami1939}. 

The range of temperatures considered in Figure~\ref{fig:NewFig1}(b), $T\lesssim T_\mathrm{h}$, is the crossover between the two regimes.  Given the opposing system-size dependence, we can expect the system size dependence to be especially small in the crossover regime.  Whether negligible or weak, the implication is that there can be no surface tension for interfaces separating domains of different metastable states.  This implication follows from
 $\log (\tau_\mathrm{MS})$ providing an upper bound to any free energy barriers that might separate distinguishable states in the metastable liquid.  In particular, two coexisting domains separated by a stable interface will interconvert on a time scale $\tau_\mathrm{int}$, and this time is proportional to $\exp(\beta \Delta F)$, where $\Delta F$ is the interfacial free energy and $\beta$ is reciprocal temperature.  To be observable, this time must be smaller than $\tau_\mathrm{MS}$.\footnote{This argument presumes that reversible two-liquid behavior requires interconversion without involvement of the crystal phase.  Were ice to be an intermediate for interconversion between a low density liquid and a high density liquid, the bound discussed herein would not apply.}

As a result, the free energy barrier for interconverting coexisting domains of low-density liquid and high-density liquid cannot be an interfacial free energy that would grow (unbounded) as a fractional power of system size.  Rather, this interfacial free energy grows at most logarithmically with system size, which implies that different forms of supercooled liquid water can be at most transient and finite in size.

To assign numbers to this bound, we can use transition state theory for a diffusive barrier, i.e., $\tau_\mathrm{int} \approx A \exp(\beta \Delta F)$, where $A \approx \tau_\mathrm{o} (\tau_\mathrm{R}/\tau_\mathrm{o})^{1-\theta}$.  Here, $\tau_\mathrm{o} \approx 1$\,ps is the characteristic microscopic time \cite{moilanen2008}, and $\theta \approx 1/4$ is the decoupling exponent relating reorganization time to diffusion \cite{swallen2003,jung2004}.  Thus, because $\tau_\mathrm{int} < \tau_\mathrm{MS}$, we arrive at the bound $\beta \Delta F < (1/4)\,\ln(\tau_\mathrm{R}/\tau_\mathrm{o}) + \ln(\tau_\mathrm{MS}/\tau_\mathrm{R})$.  Applying this formula in the regime where different liquid domains appear, i.e., $T\lesssim\Ts$, we arrive at $\beta \Delta F \lesssim 10$.  This microscopic bound to $\Delta F$ implies that away from criticality, interfaces of area larger than microscopic scales cannot be stable. Indeed, recent simulation data is consistent with this bound, with free energy estimates exhibiting a weak system size dependence before saturating at a value of $\beta \Delta F$ less than 10 \cite{Palmer2014}.

\section{Discussion}
The arguments we present herein rely on known time scales for relaxation in supercooled water and established theory relating time scales to equilibrium correlation lengths and free energy barriers.  While we focus on water, the arguments apply equally well to other supercooled liquids.  For some such materials, it is possible for $\tau_\mathrm{MS}/\tau_\mathrm{R}$ to be sufficiently large that criticality and two-liquid coexistence can be good approximations.  As noted in the Introduction, an assortment of simulation models and colloidal suspensions behave in this way. See, for example, Refs. \cite{Asherie1996,Broide1996,tenwold1997,rosenbaum1999}.

While divergent scales and two-liquid coexistence are impossible for metastable liquids in principle, nothing prohibits such behavior at conditions where a fluid is stable with respect to the crystal.   Recent work by Smallenburg \textit{et al.} \cite{Smallenburg2014} is a case in point.  That work examines a class of models with nominally tetrahedral patchy particles.  Stable liquid-liquid phase transitions do occur for some members of that class, those with parameters that suppress tetrahedral ground states in favor of BCC orderings. 
When parameters are changed to favor water-like models (i.e., locally tetrahedrally ordered), liquid-liquid behavior becomes metastable and therefore not observable beyond microscopic scales.

 \section*{Acknowledgments} 
In this work, DC has been supported by the Director, Office of Science, Office of Basic Energy Sciences, Materials Sciences and Engineering Division and Chemical Sciences, Geosciences, and Biosciences Division under the U.S. Department of Energy under Contract No. DE-AC02-05CH11231. DTL has been supported by the Princeton Center for Theoretical Science.

\section*{Appendix}
\subsection*{Time-scale separation}

\begin{figure}[t!] 
\begin{center}
\includegraphics[width=4in]{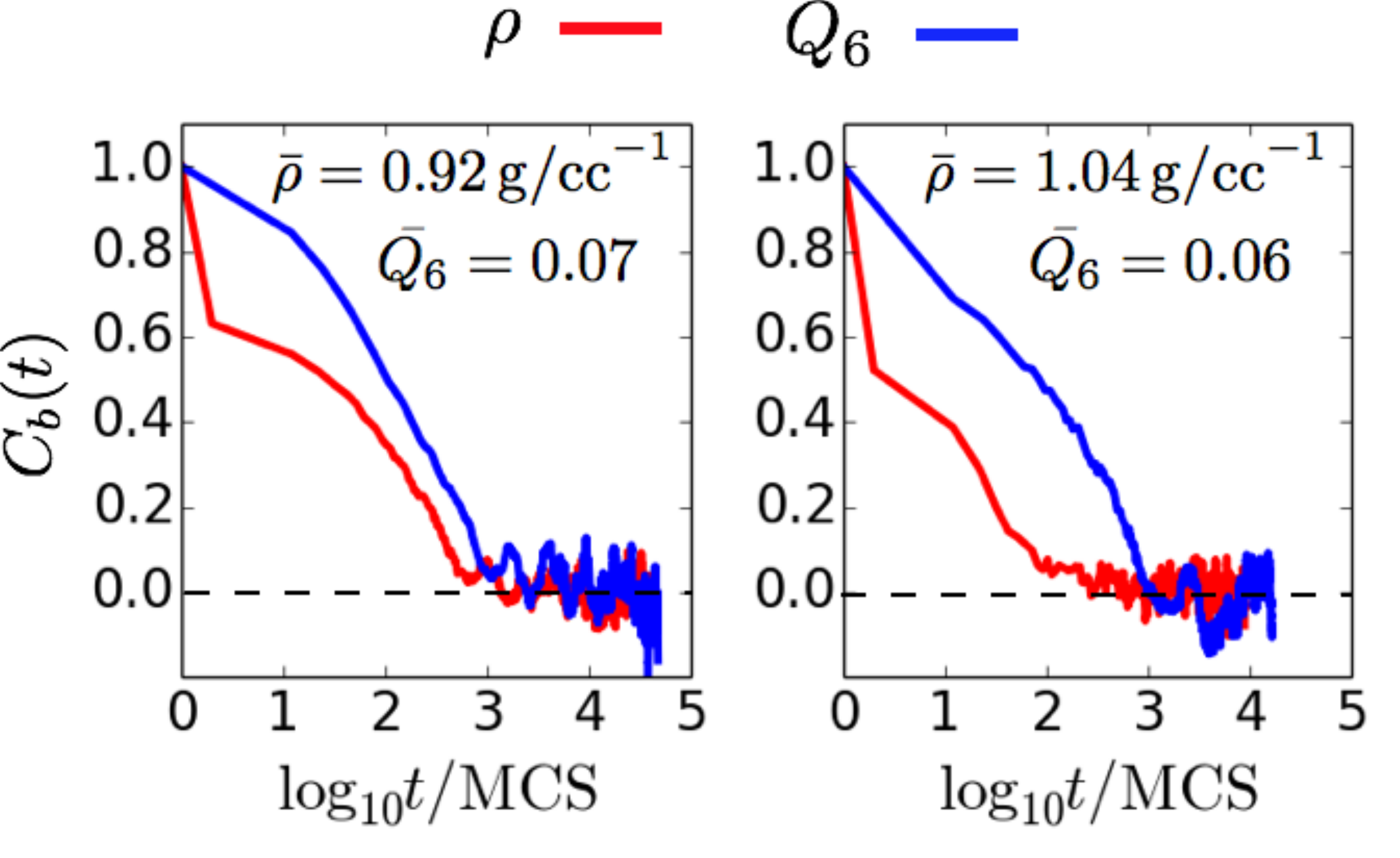}
\caption{Relaxation functions for a variant of the ST2 model of water at $p=2.2$\,kbar and $T=235$\,K as a function of Monte Carlo steps (MCS).  Red and blue lines are, respectively, the $\rho$ and $Q_6$ autocorrelation functions in two different different windows sampled during free energy calculations of Ref.~\cite{Limmer2013}.  The average value of $\rho$ and $Q_6$ in a specific window is $\bar{\rho}$ and $\bar{Q}_6$.  Random oscillations about zero at the largest times are the results of autocorrelating over finite times, typically between 50 to 100 times that for the $Q_6$-correlation function to reach 0.1 of its initial value.   Notice two (or more) step relaxation for $\rho$, and that for small $\bar{Q}_6$, the long-time relaxation times of $\rho$ increase and approach those of $Q_6$ as $\bar{\rho}$ decreases.
}
\label{fig:Fig3}
\end{center}
\end{figure} 

Separation between time scales for fluctuations in density, $\rho$, and time scales fluctuations in global crystal order, $Q_6$, are self evident because density fluctuations can be observed over periods of time where liquid water persists at supercooled conditions.  The nature of these differing scales can be examined with simulation by studying autocorrelation functions for these variables.  Doing so, however, is not straightforward because distributions of supercooled-liquid micro states are not stationary.  On the other hand, distributions are stationary for each of the constrained ensembles sampled during during the Monte Carlo calculations of the reversible-work surface, $F(\rho, Q_6)$. 

While it is difficult to assign an absolute physical time to the steps of a Monte Carlo trajectory, one may nevertheless examine relative relaxation times with the autocorrelation functions obtained within a given sampling window, $b$, as a function of time $t$ in units of Monte Carlo steps.  Each different value of $b$ coincides with a different window and thus a different value for the average $\rho$ and $Q_6$ within that window.  

Figure~\ref{fig:Fig3} shows such relaxation functions obtained in this way for $\rho$ and $Q_6$ obtained from data assembled for Ref.~\cite{Limmer2013} for the ST2 model,
\begin{equation*}
C_b(t) = \frac{\langle \delta \rho(0)\,\delta \rho(t)  \rangle_b}{\langle (\delta \rho)^2  \rangle_b} \quad \mathrm{and} \quad \frac{\langle \delta Q_6(0)\,\delta Q_6(t)  \rangle_b}{\langle (\delta Q_6)^2  \rangle_b}\,,
\end{equation*}
respectively.  Here, $\langle \cdots \rangle_b$ denote ensemble average with the biasing potential used to confine configurations to the $b$th window of $\rho$-$Q_6$ space in a free energy calculation.  The fluctuations, $\delta \rho$ and $\delta Q_6$, are deviations from their respective means in the $b$th window.

The correlation functions for the two windows considered in Fig. \ref{fig:Fig3} are representative of what we find over a swath of more than 100 separate constrained ensembles covering a broad variety of relaxation behaviors.  
The time-scale separation between $\rho$ and $Q_6$ is especially clear at the higher density characteristic of early-stage coarsening.  The time-scale separation diminishes at the longer times but remains significant at the shorter times for the lower density characteristic of ice.

\end{document}